# The exact universal exchange-only local field factor of the uniform electron gas and its near-cancellation by parallel-spin correlation near $2k_F$

Carl A. Kukkonen
33841 Mercator Isle, Dana Point, CA 92629
kukkonen@cox.net

## Abstract

I use the microscopically derived, well known, but underappreciated, exact universal expression for the exchange-only local field factor $G_x(q,\omega)$ at $\omega = 0$ to gain physical intuition about the separate exchange and correlation contributions to the static quantum Monte Carlo derived "measured" density and spin local field factors $G_+(q)$ and $G_-(q)$. I use the decomposition of $G_+$ and $G_-$ into their fundamental microscopically defined spin-resolved constituents $G_{\uparrow\uparrow}$ and $G_{\uparrow\downarrow}$. This decomposition is possible because both $G_+$ and $G_-$ are available from Monte Carlo calculations. Recognizing that $G_{\uparrow\uparrow}$ includes contributions from both exchange and correlation, $G_{\uparrow\uparrow} = G_{\uparrow\uparrow x} + G_{\uparrow\uparrow c}$ with $G_{\uparrow\uparrow x} = G_x$, allows the subtraction that reveals the parallel-spin correlation. The antiparallel-spin $G_{\uparrow\downarrow}$ depends only on correlation. The exact $G_x(q,\omega)$ is universal, independent of density, when plotted against $q/k_F$ and $\omega/\varepsilon_F$. This paper focuses on the static value at $\omega = 0$.

The density functional theory exchange and correlation kernel is directly proportional to the density local field factor through $f_{xc} = -4\pi e^2 G_+/q^2$. The local field factors define all of the response functions of the electron gas.

The static $G_x$, known in pieces for fifty years, is assembled here into one expression, evaluated exactly, and released as a short Python module. It has a large peak of 1.99 just below $2k_F$, passes through $\pi^2/6$ at $2k_F$, and tends to 1/3. Used alone in static linear response it predicts a spin density wave at $r_s = 4.96$ and a charge density wave at $r_s = 10.62$. Quantum Monte Carlo shows the gas has neither.

Near $2k_F$ the difference is striking. $G_x$ has a large peak near $2k_F$ and $G_+$ shows no comparable structure. Exchange enters $G_+$ and $G_-$ identically, so their difference, which is twice the antiparallel factor $G_{\uparrow\downarrow}$, contains no exchange at all. That difference is measured directly and is small and smooth through $2k_F$. Whatever removed the exchange-only peak is in the parallel-spin channel. There everything beyond first-order exchange, screening and correlation together, is large and negative and mirrors the exchange peak. In the measured $G_+$ the net removal is 40 to 48 percent of the peak at $r_s = 1$ to 10. The stability of the metallic gas against exchange-driven density waves is therefore a parallel-spin effect. The decomposition points to where microscopic calculations of each channel should go.

## I. Introduction

The density and spin local field factors $G_+(q,\omega)$ and $G_-(q,\omega)$ of the uniform electron gas define its response functions. They enter the effective electron-electron interaction, the dielectric function, the resistivity, the density wave instabilities, and the repulsive term μ in superconductivity [1,2]. Through $f_{xc} = -4\pi e^2 G_+/q^2$ the density factor is also the exchange and correlation kernel of density functional theory. Both are defined relative to the noninteracting Lindhard function $\chi_0$. In the density channel,

$$\chi(q,\omega) = \chi_0(q,\omega) / \{ 1 - v_q [1 - G_+(q,\omega)] \chi_0(q,\omega) \}, \qquad (1)$$

and in the spin channel,

$$\chi_s(q,\omega) = \chi_0(q,\omega) / \{ 1 + v_q G_-(q,\omega) \chi_0(q,\omega) \}, \qquad (2)$$

with $v_q = 4\pi e^2/q^2$.

Both factors are now measured by quantum Monte Carlo. Moroni, Ceperley, and Senatore (MCS) obtained $G_+$ at $r_s$ = 2, 5, and 10 in 1995 [3]. Their data are sparse. There are eleven points at $r_s$ = 2, fewer at $r_s$ = 5 and 10, and none below $k_F$. Variational diagrammatic Monte Carlo (VDMC) later gave $G_+$ at $r_s$ = 1 and 2 and $G_-$ at $r_s$ = 1 to 5, on a dense grid of 30 points with spacing $0.078k_F$, from $0.078k_F$ to $2.34k_F$ [4,5]. Kaplan and Kukkonen (KK) fitted analytic forms to both data sets, and the forms lie within the Monte Carlo error bars at every fitted density [6]. Corradini and co-workers had earlier fitted MCS alone [7]. These Monte Carlo factors are the measured truth of this paper. They are accurate enough that subtracting an exact exchange part from them is meaningful, and the one place where the data sets disagree is stated where it matters. Finite temperature is treated by path integral Monte Carlo [8]. This paper concerns the ground state. MCS found $f_{xc}$ nearly constant below $2k_F$, which is the adiabatic local density approximation (ALDA), and called that intriguing. They looked for the $2k_F$ hump of the approximate theories and found no evidence of it. Both observations are taken up here.

The exchange-only local field factor $G_x(q,0)$ is exact and universal. It is a single curve, the same at every density. In exchange-only theory $G_+ = G_- = G_x$, because the Pauli principle acts only between like spins. The function has been known in pieces since 1970. Its small-q and large-q limits, its ratio form, its peak near $2k_F$, its universality, and its reduction to elementary quadratures were each established by others. Table I in Section II lists the pieces and who found them. This paper assembles them into one expression, Eq. (6), evaluates it exactly at every wavevector including $2k_F$ itself, where every earlier quadrature failed, and releases it as a short Python module that returns $G_x$ for any q. The value at $2k_F$ is $\pi^2/6$. I have not found it stated before.

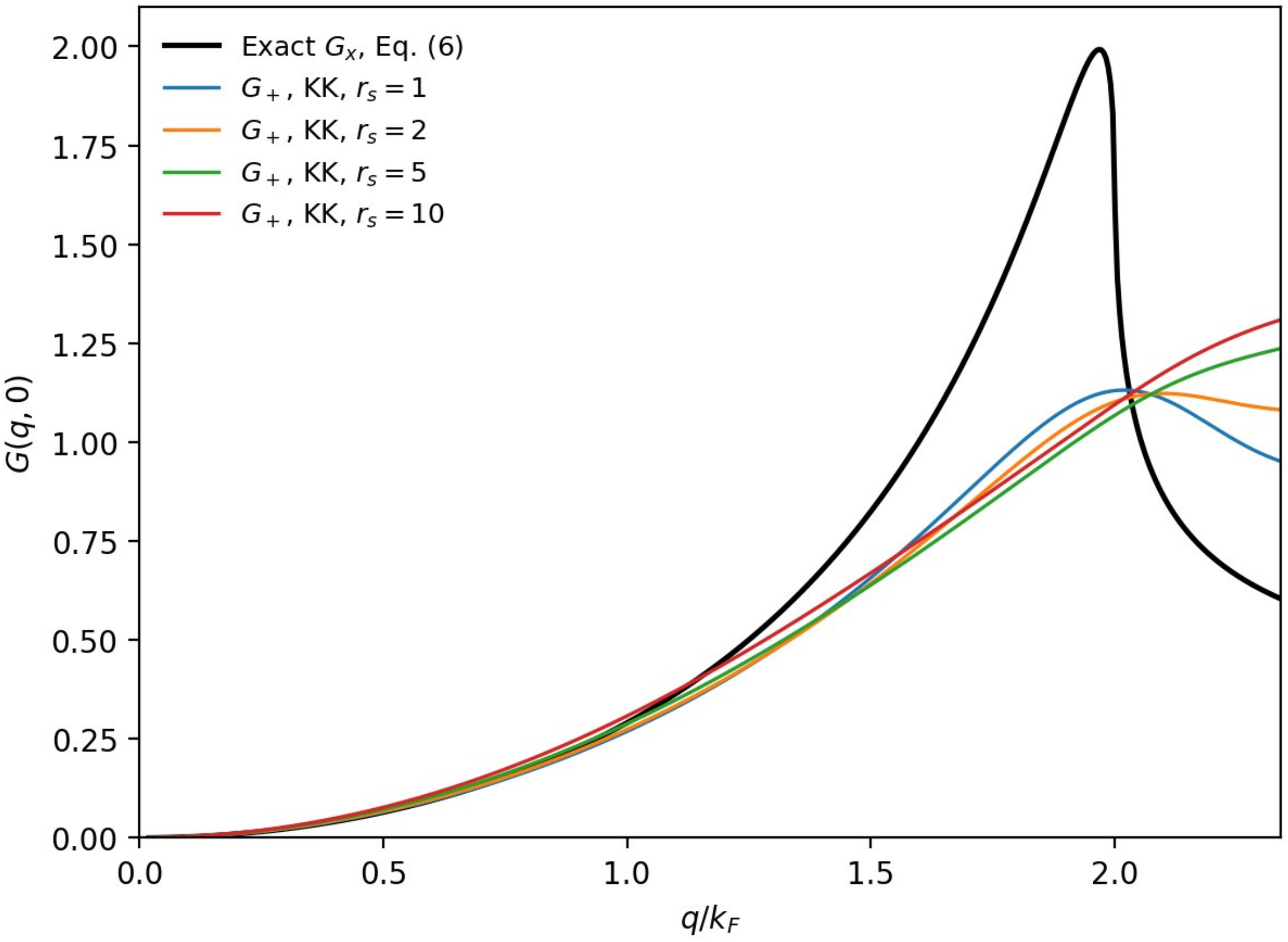

Fig. 1. The exact universal static exchange-only local field factor $G_x(q,0)$, black, Eq. (6), compared with the full $G_+(q,0)$ at $r_s$ = 1, 2, 5, and 10 from the Kaplan-Kukkonen representation [6], the densities where it is anchored to Monte Carlo data.

Figure 1 shows the two objects. Exchange alone produces a peak near $2k_F$ of height 1.99, roughly 1.5 to 1.8 times the height of the full factor at the densities shown, while the full $G_+$ curves are close to one another, smooth, and free of $2k_F$ structure. The gap between the black curve and the colored ones is the subject of this paper.

Used alone in static linear response, $G_x$ predicts a spin density wave at $r_s = 4.96$ and a charge density wave at $r_s = 10.62$. The three-dimensional electron gas has neither. Quantum Monte Carlo finds the paramagnetic fluid stable far beyond both densities and finds no significant $2k_F$ structure in $G_+$ or $G_-$ [3,4,9]. Overhauser proved that within Hartree-Fock a density wave lies below the uniform state at every density [10]. Whether correlation removes it was argued both ways for decades. Quantum Monte Carlo settled it.

So screening and correlation must remove the exchange peak, and the exact $G_x$ shows how. Exchange enters $G_+$ and $G_-$ identically, and their difference $G_+ - G_- = 2G_{\uparrow\downarrow}$ contains no exchange at all. Subtracting $G_x$ from the Monte Carlo $G_+$ and $G_-$ therefore separates parallel-spin correlation from antiparallel-spin correlation exactly. Throughout, correlation means everything in the local field factor beyond first-order exchange. Screened exchange and higher-order exchange are counted as correlation. Section II gives the exact $G_x$ and its history. Section III derives the exchange-only instabilities. Section IV performs the subtraction on the Monte Carlo data. Section V benchmarks the two published exchange-only kernels against the exact curve. Section VI states the consequences. Appendix A gives the Engel-Vosko function. Appendix B collects the numerical values used in the text, each with its source.

## II. The exact universal exchange-only local field factor

This section supplies the tool. It defines $G_x$, writes it as one expression, gives its history, its exact values, and the physics of its peak, and shows it against the two earlier tabulations. Everything after this section uses Eq. (6) and nothing else about exchange.

The local field factor is related to the proper polarizability Π by

$$G(q,\omega) = [\chi_0(q,\omega) - \Pi(q,\omega)] / [v_q\, \Pi(q,\omega)\, \chi_0(q,\omega)]. \qquad (3)$$

The proper polarizability expands in powers of $e^2$ as $\Pi = \chi_0 + \Pi^1 + \ldots$, where $\Pi^1$ is the first-order exchange term. Substituting this into Eq. (3) term by term, the numerator becomes $\chi_0 - \Pi = -\Pi^1 - \ldots$, the denominator becomes $v_q(\chi_0 + \Pi^1 + \ldots)\chi_0$, and

$$G(q,0) = -\Pi^1 / [v_q\, \chi_0^2\, (1 + \Pi^1/\chi_0 + \ldots)].$$

The ratio $-\Pi^1/v_q\chi_0^2$ is of order one, because $\Pi^1$ and $v_q$ are each of order $e^2$. The factor in parentheses is 1 plus corrections of order $e^2$. Truncating consistently at first order means dropping those corrections, which gives the exchange-only local field factor

$$G_x(q,0) = -\Pi^1(q,0) / [v_q\, \chi_0(q,0)^2]. \qquad (4)$$

The truncation matters. $G_x$ so defined is exactly first order in $e^2$ and therefore universal. Keeping the parenthesis instead, that is, inserting the first-order polarizability into Eq. (3) unexpanded, retains the higher-order corrections, and spurious powers of $r_s$ return. The result is neither first order nor universal [11]. Qian writes that untruncated form explicitly, attributes it to BDL, and notes that it is not the local field factor with the exact exchange contribution [12, II, Eq. (1)]. Equation (4) is Eq. (5.229) of Giuliani and Vignale [1], it is the object whose universality Brosens, Devreese, and Lemmens (BDL, whose author order varies among their papers) proved [13], and it is the static limit of the exact-exchange kernel of time-dependent density functional theory [14,15,16].

It is convenient to work with the dimensionless function $I(Q) = (\pi^3/m^2e^2)\Pi^1(q,0)$, where $Q = q/2k_F$. Engel and Vosko (EV) give $I(Q)$ as an explicit expression in logarithms and two elementary one-dimensional integrals [17], reproduced in

Appendix A. The static Lindhard function enters through $\chi_0(q,0) = -(mk_F/\pi^2)L(Q)$, with

$$L(Q) = 1/2 + [(1 - Q^2)/4Q] \ln|(1 + Q)/(1 - Q)|, \qquad (5)$$

and Eq. (4) becomes

$$G_x(q,0) = -I(Q)\, Q^2 / L(Q)^2, \quad Q = q/2k_F. \qquad (6)$$

Equation (6) is the exact static exchange-only local field factor of the uniform electron gas. It is an explicit expression evaluated by two elementary quadratures and given in Appendix A. It is released as a short Python module. Below $2k_F$ Glasser gives it with no integral at all [18]. Everywhere Qian gives it as a convergent series [12]. Only the Engel-Vosko form is evaluated in this paper. The density has disappeared from the right-hand side. There is one curve, not a family of curves, and it is the same at every electron density once q is measured in units of $k_F$.

The exchange-only local field factor has been studied for more than fifty years. Table I lists what is known about it and who established each piece. Each entry built on the ones before. Everything except the last row is due to others.

Table I. The exchange-only local field factor, static and dynamic. What is known, who established it, and when. The entries before 1990 are as reported in Refs. [19], [13], and [20]. The reader will find further references there and in Giuliani and Vignale, Chapter 5 [1].

| **What is known** | **Who and when** | **Note** |
|---|---|---|
| The local field factor G(q) as a correction to the Lindhard response, with large-q value 1/2 | Hubbard 1957 [21] | The 1/2 is the Hubbard defect discussed in Section V. |
| First-order exchange correction to the static screening function. Small-q coefficient 1/4. Large-q limit. | Geldart and Taylor 1970 [22]. Written as G → 1/3 by Geldart, Richard, and Rasolt 1972 [23]. Restated by Zhu and Overhauser 1984 [24]. | The 1/3 predates the 1977 papers. |
| The ratio form $G = P_0/Q_0$. Density independence. A peak near $2k_F$, 0.81 at $1.95k_F$. | Toigo and Woodruff 1970, 1971 [25,26] | Their decoupling averages over frequency. The peak is 40 percent of the true height and the large-q limit is 2/3, corrected in Ref. [23]. Not the first-order function, so not plotted here. |
| The formal expression for the first-order dynamic G(q,ω), whose ω = 0 limit is Eq. (4) | Rajagopal 1972 [27]. Brosens, Lemmens, and Devreese 1976 [28]. Tripathy and Mandal 1977 [19]. | Three independent routes to one expression. TM was received in October 1975 and published in July 1977. BDL and TM do not cite each other. |
| Universality. $G_x$ depends only on $q/k_F$ and $\omega/\varepsilon_F$, static and dynamic. | Brosens, Lemmens, and Devreese 1977 [13] | The theorem. TM remark only that their static G is the same for all $r_s$. |
| First tabulations and plots of the static function. Peak near $1.97k_F$. Accuracy about 1.5 percent. | Tripathy and Mandal 1977 [19]. Brosens, Lemmens, and Devreese 1977 [29] as reported in Ref. [30]. Brosens, Devreese, | Every quadrature fails at exactly $2k_F$. TM give 1.751 and BDL 1.531 against $\pi^2/6$ = 1.645. See Fig. 2 and Table III. |
| | and Lemmens 1980 [30]. | |
| g(0) = 1/2 from the large-q limit by both the Niklasson and the Kimball routes | Brosens, Lemmens, and Devreese 1977 [13]. Niklasson 1974 [31]. Kimball 1973 [32]. | The exchange-only contact value. |
| The universal static curve plotted against approximate kernels. Conjecture that correlation washes out the peak. | Holas, Aravind, and Singwi 1979 [11] | Also the first evaluation at arbitrary (q,ω). |
| Higher small-q coefficients of $\Pi^1$ | Kleinman and Lee 1988, Chevary and Vosko 1990 [33]. Svendsen and von Barth 1996 [34]. | |
| Reduction of $\Pi^1(q,0)$ to elementary one-dimensional quadratures. $I(1) = -\pi^2/24$. One diagram by an ansatz. | Engel and Vosko 1990 [17] | The form used here, Eq. (A1). The large-q term is Kleinman 1984 [35]. |
| I(Q) in dilogarithms and trilogarithms with no remaining integral for $q \le 2k_F$ | Glasser 1995 [18] | Also a proof of the Engel-Vosko form for $q \le 2k_F$. |
| The Engel-Vosko ansatz proved at all q. A convergent series. The $\ln^3$ expansion at $2k_F$. | Qian 2015, 2016 [12] | Qian evaluates $\Pi^1$ and does not form G. He notes that the untruncated form is not the local field factor with the exact exchange contribution. |
| The TDDFT exact-exchange kernel is the same object. RPAx instability at $r_s \approx 10.6$. | Görling 1998 [16]. Hellgren and von Barth 2008 [14]. Colonna, Hellgren, and de Gironcoli 2014 [15]. | The BDL 1980 minimum rediscovered by a different route. |
| The assembled expression (6). $\pi^2/6$ at $2k_F$. The exact peak 1.9924 at $1.9685k_F$. Threshold curves in closed form. The Python module. The spin-resolved subtraction against QMC. | This paper | |

Three values of Eq. (6) are exact and follow from the special values of I(Q) given by EV. At small q, $I(0) = -1$ and $L(0) = 1$ give $G_x \to (1/4)(q/k_F)^2$, the coefficient fixed by the Hartree-Fock compressibility sum rule. At exactly $q = 2k_F$, $|I(1)| = \pi^2/24$ and $L(1) = 1/2$ give

$$G_x(2k_F) = \pi^2/6 = 1.6449\ldots , \qquad (7)$$

and at large q the leading term $I(Q) \to -(1/27)Q^{-6}$ combines with $L(Q) \to 1/(3Q^2)$ to give

$$G_x(q \to \infty) = 1/3. \qquad (8)$$

Between the small-q rise and the large-q limit the function has a smooth maximum of 1.9924 at $q = 1.9685k_F$, $0.031k_F$ below $2k_F$. The maximum is finite but the slope at $2k_F$ is not. EV showed that the slope of $\Pi^1(q,0)$ diverges as $(\ln|q - 2k_F|)^3$ there, a stronger non-analyticity than the single logarithm of the Lindhard function, and $G_x$ inherits it on both sides [17]. Table II shows how the two factors of Eq. (6) produce the shape.

Table II. The two factors of Eq. (6) at the wavevectors that fix its shape.

| $Q = q/2k_F$ | $-I(Q)$ | $L(Q)$ | $G_x = -I\,Q^2/L^2$ |
|---|---|---|---|
| 0 | 1 | 1 | $Q^2 \to 0$ |
| 0.984 (numerical) | 0.596 | 0.538 | 1.9924, the maximum |
| 1 | $\pi^2/24$ | 1/2 | $\pi^2/6$ |
| $\to \infty$ | $1/(27Q^6)$ | $1/(3Q^2)$ | 1/3 |

The peak has a simple origin in the two factors. Below $2k_F$ the Lindhard denominator $L^2$ falls faster than the exchange numerator I, so the ratio climbs. Above $2k_F$ the numerator collapses as $Q^{-6}$ and the ratio falls to 1/3. The physics behind the arithmetic is Fermi-surface kinematics. L(Q) is the phase space for particle-hole pairs of zero energy, and it carries the Kohn logarithm at $2k_F$, the largest momentum that connects two points on the Fermi surface. I(Q) is the same phase space with one bare Coulomb line inserted between the particle and the hole. The $1/k^2$ weight of that line at small momentum transfer turns the single logarithm into a cubed one. In EV's decomposition the cubed logarithm lives entirely in the vertex graph, and the self-energy graphs carry only a squared logarithm [17, Eqs. (9), (18), (25)]. The peak is therefore Fermi-surface kinematics amplified by the unscreened Coulomb line. Screening that line is what should tame it, and Section VI returns to this. EV noted one consequence of the undamped cusp. It would change the screening cloud around an impurity from $\cos(2k_Fr)/r^3$ to $\cos(2k_Fr)(\ln 4k_Fr)^2/r^3$ and would sharpen Kohn anomalies [17].

The large-q limit carries more information than its value. Niklasson established the exact relation $\lim_{q\to\infty} G(q,0) = 2[1 - g(0)]/3$ between that limit and the pair correlation function at zero separation [31], see also Kimball [32], Zhu and Overhauser [24], and Ichimaru [36]. The fully correlated factor does not have a large-q limit. It grows as $C(r_s)(q/k_F)^2$, where C is a density-dependent coefficient computed by Zhu and Overhauser [24]. That growth comes from the interaction-induced change in the occupation numbers, a kinetic effect that vanishes at first order in $e^2$, so the limit exists for $G_x$. Substituting the exchange-only limit 1/3 into the Niklasson relation gives g(0) = 1/2. That is zero contact probability for parallel spins by the Pauli principle and one for antiparallel spins, as required of any exchange-only theory built on a single Slater determinant. The limit 1/3 is Geldart and Taylor's [22,23]. Tripathy and Mandal obtained it analytically [19]. Its reading as g(0) = 1/2 through the contact relation is BDL's [13, Eqs. (24)-(25)]. Zhu and Overhauser state the limit outright [24, Eq. (40)]. Two consequences follow. An exchange-only kernel is obliged to reproduce 1/3, whatever framework produced it. And because the fully correlated factor grows as $q^2$ while the exchange-only one saturates, comparisons between $G_x$ and $G_+$ are made below for $q \le 2k_F$, where the two are of the same character.

Two remarks on conventions. Niklasson defines $\chi^0$ with the interacting occupation numbers and Zhu and Overhauser with the free Lindhard function. For an exchange-only theory built on a single Slater determinant they are the same, since $n_k = \theta(k_F - k)$ exactly. And EV note that $\Pi^1$ determines the exchange contribution to the spin response as well, so $G_{+x} = G_{-x} = G_x$. The Pauli principle acts only between like spins, and this identity is what Section IV rests on.

The first-order kernel is exact at every frequency and universal in $q/k_F$ and $\omega/\varepsilon_F$ together, as BDL proved [13]. What is particular to $\omega = 0$ is not the exactness but the explicit expression. Equation (6) is the static section of a universal surface. The dynamic kernel is the subject of a companion paper.

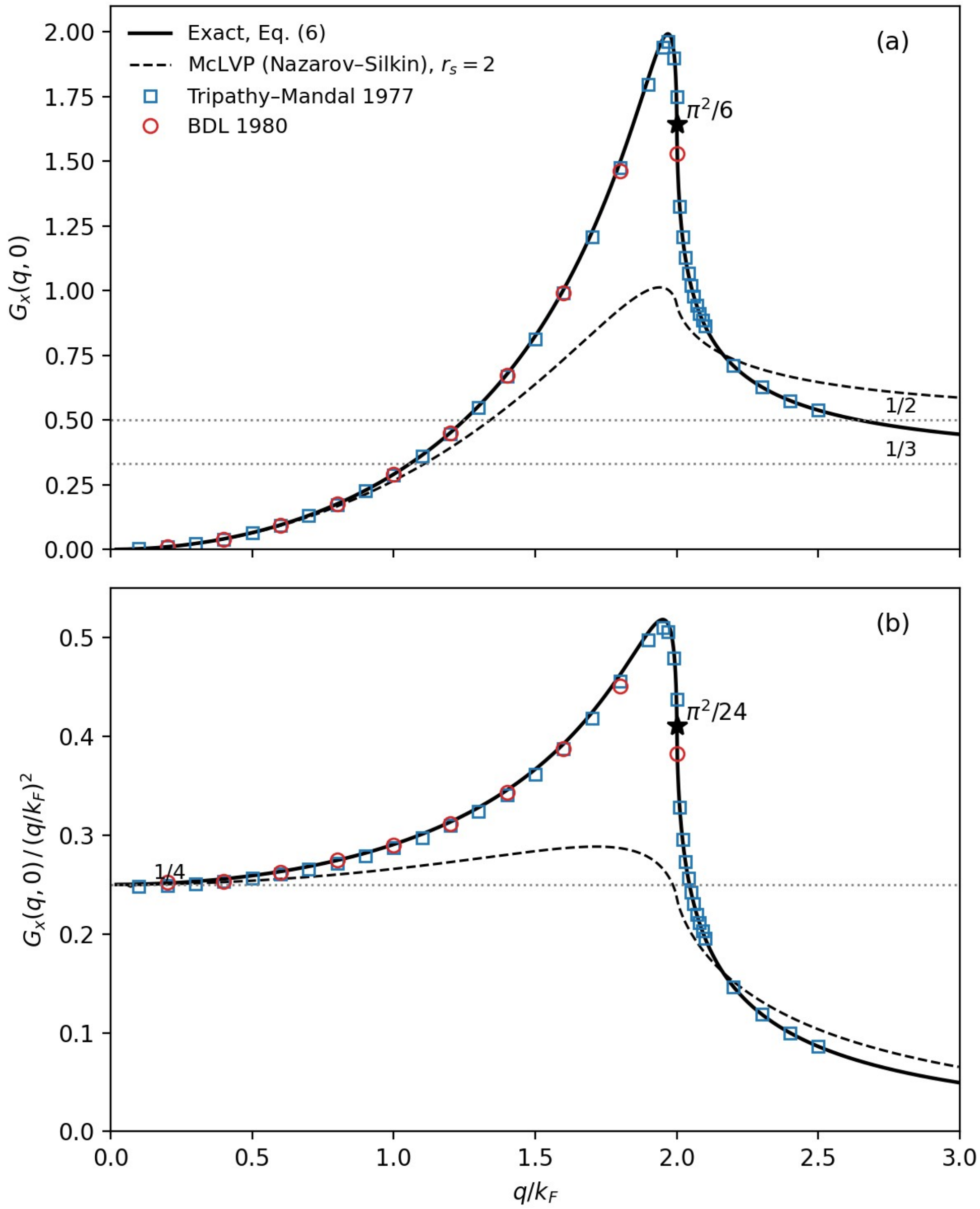

Fig. 2. (a) The exact static exchange-only local field factor $G_x(q,0)$, Eq. (6), solid line, evaluated with the Python module ev_gx.py deposited with this paper. Open squares are the 1977 tabulation of Tripathy and Mandal, their Table I [19]. Open

circles are the 1980 tabulation of Brosens, Devreese, and Lemmens, the $\nu = 0$ row of their Table I [30], $\nu$ being their dimensionless frequency. The dashed line is the McLachlan-variational kernel of Nazarov and Silkin at $r_s = 2$ [37]. The star marks $\pi^2/6$ at $2k_F$. Dotted lines mark the exact large-q limit 1/3 and the McLVP limit 1/2. (b) The same quantities divided by $(q/k_F)^2$, which is the exchange kernel in dimensionless form.

Figure 2 shows the curve against the two earlier tabulations and the McLVP kernel. Both 1977 and 1980 tabulations lie on the exact curve everywhere except at $2k_F$ itself, where one is high and one is low and $\pi^2/6$ sits between them. That is what a quadrature not adapted to a cubed logarithm does, and it is why the analytic value is useful. Nazarov and Silkin supplied McLVP data at $r_s = 1, 2, 5$, and 10, and in these units the four sets collapse onto one curve, so their kernel shares the universality. Its departure from the exact curve near $2k_F$ is discussed in Section V. The lower panel bears on the MCS observation quoted in the Introduction. A constant $f_{xc}$ below $2k_F$ is the assumption of the ALDA, and MCS found the measured kernel nearly constant there. The exchange part alone is not constant. It rises by a factor $\pi^2/6$, 65 percent, from 1/4 at $q = 0$ to $\pi^2/24$ at $2k_F$. The observed constancy must therefore come from a cancellation, and Section IV shows what cancels.

The Python module ev_gx.py evaluates Eq. (6) from Appendix A alone. It takes $q/k_F$ and returns $G_x$, $I(Q)$, and $L(Q)$. It reproduces $I(0) = -1$ and $I(1) = -\pi^2/24$ to machine precision. Every exact number in this paper was produced by it, and it is deposited with the paper [38]. Anyone who needs the exact static exchange kernel, or a benchmark for an approximate one, can call it.

## III. Exchange-only instabilities of the uniform electron gas

This section asks what the gas would do if exchange were the only correction to the noninteracting response, and answers with two predicted instabilities. Neither occurs in the real gas. The gap between prediction and reality is what Section IV measures.

Instabilities are read from Eqs. (1) and (2). An external potential $V_{ext}(q)$ induces a density $n(q) = \chi(q,0)V_{ext}(q)$, and a magnetic field that couples to the spins induces a spin density $m(q) = \chi_s(q,0)B_{ext}(q)$. An instability is the statement that a response denominator vanishes at some wavevector. The induced modulation then survives as the applied field goes to zero. If the density denominator in Eq. (1) vanishes the result is a static charge density wave (CDW). If the spin denominator in Eq. (2) vanishes the result is a spin density wave (SDW).

Start at $q = 0$, where the two channels are governed by exact sum rules. The $q \to 0$ limits of $\chi$ and $\chi_s$ are fixed by the compressibility $\kappa$ and the spin susceptibility $\chi_\sigma$, which are second derivatives of the ground-state energy. The exchange energy is known analytically, so the exchange-only corrections to both are exact, and they are the same,

$$\kappa_0/\kappa = \chi_{0\sigma}/\chi_\sigma = 1 - r_s/(\pi\alpha), \quad \alpha = (9\pi/4)^{1/3}, \qquad (9)$$

where the subscript zero marks the noninteracting values and $\pi\alpha = 6.03$. The two channels are identical at $q = 0$ because $G_+$ and $G_-$ share the same small-q coefficient, Section II. What a vanishing denominator means, however, is entirely different in the two channels, and the difference is the uniform background.

Flipping spins at fixed total density costs no electrostatic energy. So Eq. (9) is the whole story for spin at long wavelength. The enhancement $\chi_\sigma/\chi_{0\sigma}$ diverges at $r_s = \pi\alpha = 6.03$, and exchange-only theory predicts a ferromagnetic instability there. Bloch obtained the finite-amplitude version of the same physics in 1929, full polarization beyond $r_s = 5.45$ [39]. Changing the total electron density is different. The rigid uniform background makes a net charge accumulation infinitely costly at long wavelength. Formally, the charge denominator in Eq. (1) is $1 + v_q(1 - G_+)|\chi_0|$, written here with $\chi_0 = -|\chi_0|$, and the Hartree term $v_q|\chi_0|$ diverges as $q \to 0$. The ratio $\kappa_0/\kappa$ passes through zero at the same $r_s = 6.03$, so the compressibility diverges there and is negative beyond it, but no long-wavelength instability follows. The background absorbs it.

At finite q the protection fades and the exchange peak takes over. The Hartree cost falls as $q^2$, and near $2k_F$ the exchange enhancement is at its maximum. The charge denominator can vanish only if the effective interaction $v_q(1 - G_+)$ turns attractive, which requires $G_+ > 1$. The spin denominator $1 - v_q G_-|\chi_0|$ carries no Hartree term, so any positive $G_-$ works against it at every wavevector. An SDW is easy. A CDW is hard.

Each condition can be solved for the density because $G_x$ is universal. With $Q = q/2k_F$ and $k_F = \alpha/r_s$, the Coulomb and Lindhard factors combine to

$$v_q|\chi_0(q,0)| = r_s L(Q)/(\pi\alpha Q^2). \qquad (10)$$

This is linear in $r_s$, while $G_x(Q)$ carries no density dependence. At each wavevector the instability condition is a linear equation in one unknown. Setting $G_+ = G_- = G_x$ gives the two threshold densities

$$r_s^{SDW}(q) = \pi\alpha Q^2/[G_x(Q)L(Q)], \quad r_s^{CDW}(q) = \pi\alpha Q^2/[(G_x(Q) - 1)L(Q)], \qquad (11)$$

plotted in Fig. 3. At a given wavevector the uniform gas is stable below the curve and unstable above it. Instability at any single wavevector destroys the uniform state, so the physical number in each case is the minimum of the curve, together with the wavevector of the wave that appears first.

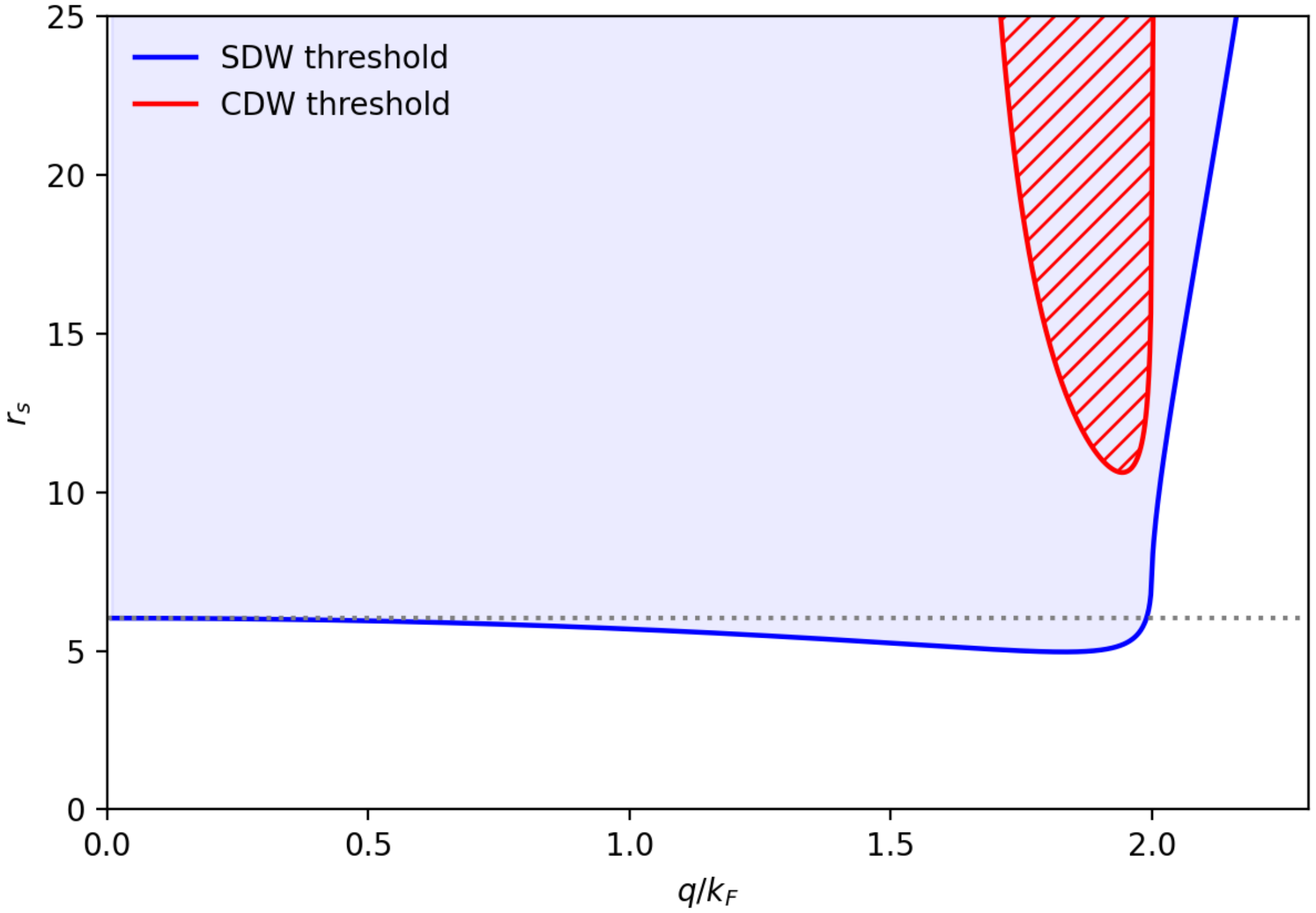

Fig. 3. Exchange-only stability boundaries, Eq. (11), from the exact $G_x$ of Eq. (6). At a given wavevector the uniform gas is stable below a curve and unstable above it. Shading marks the region unstable to a spin density wave. Hatching marks the region unstable to a charge density wave, which lies inside it. Blue, the SDW threshold, with onset at $r_s = 4.96$ and $q = 1.83k_F$. The dotted line is the long-wavelength value $\pi\alpha = 6.03$ of Eq. (9). Red, the CDW threshold, which exists only where $G_x > 1$, with onset at $r_s = 10.62$ and $q = 1.944k_F$.

The SDW threshold exists at every wavevector because $G_x > 0$ everywhere, and it is nearly flat. It falls from $\pi\alpha = 6.03$ at $q \to 0$ to a minimum of $r_s = 4.96$ at $q = 1.83k_F$ and climbs steeply beyond $2k_F$. The flatness has a simple origin. Because $G_x = -I(Q)Q^2/L(Q)^2$, two factors of L cancel and the spin threshold reduces to

$$r_s^{SDW}(q) = \pi\alpha\, L(Q)/|I(Q)|. \qquad (12)$$

L(Q) is the static polarizability of the free gas and I(Q) is the same polarizability with one exchange interaction inserted, so both decline together as the zero-energy particle-hole phase space shrinks, and their ratio moves little. The $2k_F$ exchange peak lowers the long-wavelength threshold by only 18 percent, from 6.03 to 4.96, but it moves the first unstable wavevector from $q = 0$ to $1.83k_F$.

The CDW threshold is different in every way. It exists only where $G_x > 1$, the narrow window $1.60k_F < q < 2.05k_F$, and its minimum is $r_s = 10.62$ at $q = 1.944k_F$. It is governed by $G_x - 1$, not by $G_x$. Bringing it down to 10.62 takes $G_x = 1.957$, essentially the full height of the exchange maximum. The charge density wave is driven by how far the peak rises above unity.

BDL located "a minimum in these critical values of $r_s$ as a function of q ... for $r_s \approx 10.6$ near $q = 1.94k_F$" [30], and Colonna, Hellgren, and de Gironcoli, solving the RPAx Dyson equation with the first-order kernel, found $r_s > 10.6$ at $q \approx 2k_F$ [15]. Agreement across three calculations, two numerical and one analytic, is a check on Eq. (6).

Two caveats bound these thresholds. They are linear-response conditions built from a static local kernel. Overhauser's theorem is a finite-amplitude variational result with the full nonlocal exchange interaction, and it gives density waves at every density in Hartree-Fock [10], confirmed by modern unrestricted Hartree-Fock calculations [40]. No local static kernel reproduces that. And the RPAx pole is not an exact first-order statement. Colonna and co-workers call the divergence an artifact of truncating the kernel expansion at first order [15], and alternative resummations remove it [41,42]. What survives every construction is the input supplied exactly here, the strong exchange enhancement of the $2k_F$ channel.

None of this happens in the real electron gas. Quantum Monte Carlo finds no ferromagnet, no density wave of either kind at these densities, and places the Wigner crystal near $r_s \approx 100$ [9,43]. Something suppresses all of it, and only correlation is available to do the suppressing. Replacing $G_x$ by the interacting KK factors in Eq. (11) confirms it within their fitted range. For $r_s \leq 10$ the charge condition has no solution. The spin condition has no solution at the anchored densities $r_s \leq 5$, and none through the KK extrapolation of $G_-$ up to $r_s = 10$. The exchange-only SDW at 4.96 falls inside the anchored range and does not occur. The ferromagnet at 6.03 relies on the extrapolated $G_-$, one unit beyond its last anchor. The exchange-only CDW at $r_s = 10.62$ lies just outside the $G_+$ fitted range, so the CDW statement is a near miss rather than a covered test. Section IV measures the suppression on the Monte Carlo data.

## IV. Beyond exact exchange: the effect of the Coulomb interaction and higher-order terms (correlation)

The purpose of this section is first to convince the reader that the quantum Monte Carlo calculations of the density and spin local field factors are accurate enough to stand for the true $G_+$ and $G_-$. Their sum and difference can then be taken to reveal the fundamental spin-resolved constituents $G_{\uparrow\uparrow}$ and $G_{\uparrow\downarrow}$. The antiparallel $G_{\uparrow\downarrow}$ contains no exchange and is smooth and featureless compared with $G_x$. The parallel $G_{\uparrow\uparrow}$ contains both exchange and correlation and is also smooth near $2k_F$. Subtracting the exact $G_x$ from it reveals what the Coulomb interaction and all higher-order terms must do to cancel the dramatic structure of $G_x$. These two observations reveal the spin-resolved structure that correlation must have to produce the Monte Carlo results. This is the central and simple result of this paper.

The density and spin local field factors are combinations of the parallel-spin and antiparallel-spin factors, $G_\pm = G_{\uparrow\uparrow} \pm G_{\uparrow\downarrow}$ [44,6]. Since exchange acts only between like spins, $G_{\uparrow\uparrow} = G_x + G_{\uparrow\uparrow c}$ and $G_{\uparrow\downarrow}$ is pure correlation. Therefore $G_+ - G_- = 2G_{\uparrow\downarrow}$ contains no exchange at all and $G_+ + G_- = 2G_x + 2G_{\uparrow\uparrow c}$. On the VDMC grid at $r_s = 1$ and 2, where both $G_+$ and $G_-$ exist, the decomposition needs no representation.

The data behind the KK forms are as follows [6]. $G_-$ was fitted to VDMC data on the same 30-point grid at $r_s = 1$ to 5. $G_+$ was fitted to VDMC data at $r_s = 1$ and 2 on the same grid, and to the MCS data at $r_s = 5$ and 10 over $1 < q/k_F < 4.25$. The forms are constrained by exact limits at both ends. $G_+$ is therefore anchored to data at $r_s = 1$, 2, 5, and 10 and interpolated between anchored densities. $G_-$ is anchored at $r_s = 1$ to 5, interpolated between, and extrapolated above. The data with error bars and the quality of the KK fit are shown in Fig. 4.

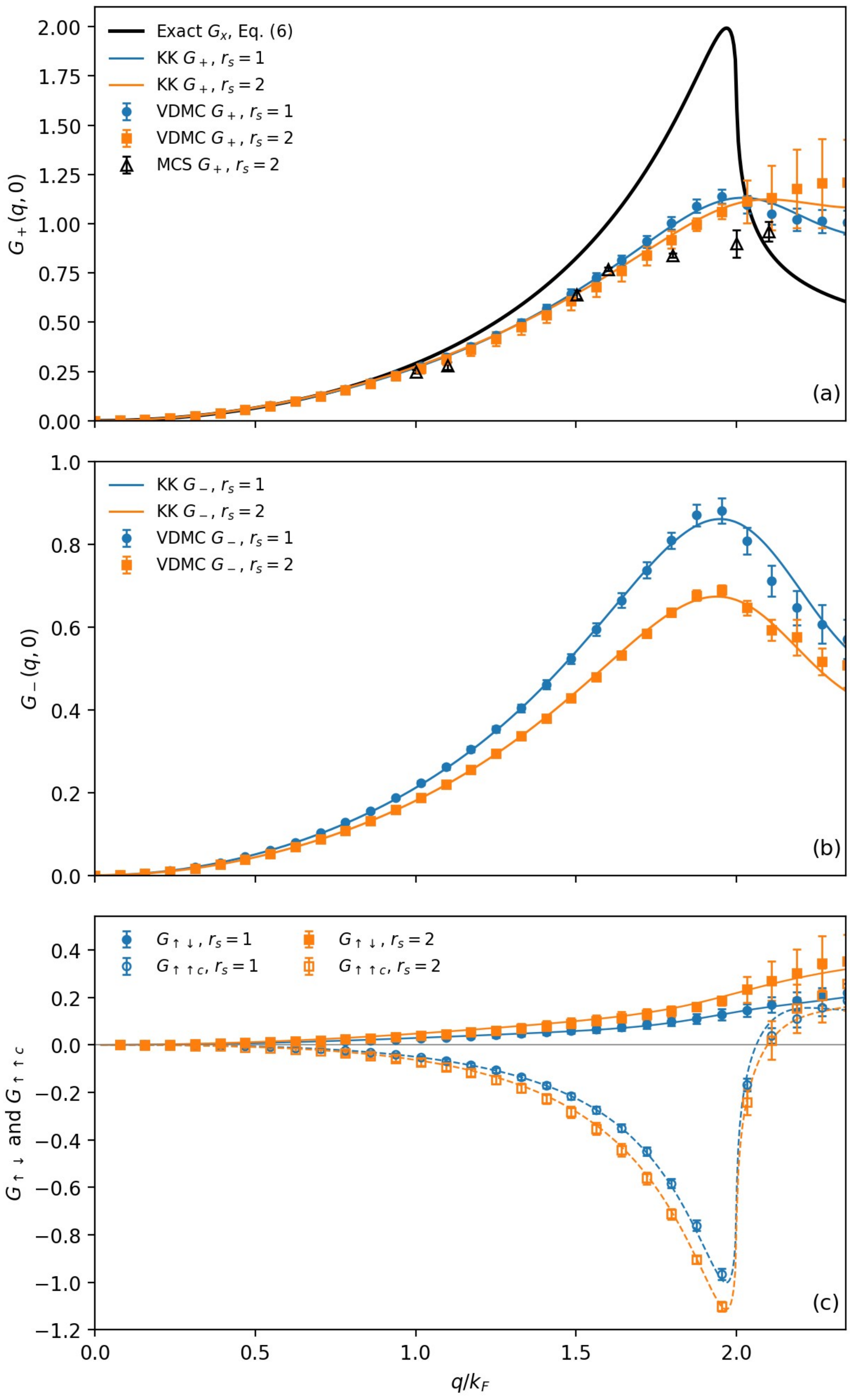

Exact $G_x$, Eq. (6)
KK $G_+$, $r_s = 1$
KK $G_+$, $r_s = 2$
VDMC $G_+$, $r_s = 1$
VDMC $G_+$, $r_s = 2$
MCS $G_+$, $r_s = 2$
$G_+(q, 0)$
(a)
KK $G_-$, $r_s = 1$
KK $G_-$, $r_s = 2$
VDMC $G_-$, $r_s = 1$
VDMC $G_-$, $r_s = 2$
$G_-(q, 0)$
(b)
$G_{\uparrow\downarrow}$, $r_s = 1$
$G_{\uparrow\uparrow c}$, $r_s = 1$
$G_{\uparrow\downarrow}$, $r_s = 2$
$G_{\uparrow\uparrow c}$, $r_s = 2$
$G_{\uparrow\downarrow}$ and $G_{\uparrow\uparrow c}$
(c)
$q/k_F$

Fig. 4. Quantum Monte Carlo data points, Kaplan-Kukkonen fits, and the exact exchange curve. (a) The exact $G_x(q,0)$, Eq. (6), solid black. VDMC $G_+$ at $r_s = 1$ and 2 with error bars [4,5]. MCS $G_+$ at $r_s = 2$ [3]. KK representations at $r_s = 1$ and 2 [6]. (b) VDMC $G_-$ at $r_s = 1$ and 2 with error bars [4,5] and the KK representations. (c) The spin-resolved pieces formed from the VDMC points at the same two densities. Filled symbols are the antiparallel factor $G_{\uparrow\downarrow} = (G_+ - G_-)/2$. Open symbols are the parallel-spin correlation $G_{\uparrow\uparrow c} = (G_+ + G_-)/2 - G_x$. Errors are propagated from the Monte Carlo errors. Solid and dashed curves are the same quantities from the KK forms.

Figure 4 makes the accuracy case and the measurement together, with no fit involved in the data points. In Fig. 4(a) the KK curves pass through every VDMC point, and the VDMC and MCS sets agree below $1.9k_F$. $G_+$ tracks $G_x$ to about $0.8k_F$ and then falls away, and near $2k_F$ it is roughly half of $G_x$. At the VDMC point closest to the exchange maximum, $1.953k_F$, correlation removes 42 percent of exchange at $r_s = 1$ and 46 percent at $r_s = 2$. At the MCS point $1.81k_F$ it removes 40 to 45 percent at $r_s = 2$, 5, and 10. So the removal is roughly half of the exact exchange value near its maximum, at four densities, from two independent Monte Carlo methods. One grid step later, at $2.031k_F$, exchange and the full factor are equal within the errors, and beyond that $G_+$ exceeds $G_x$. The values, their statistical errors, the corresponding numbers through the KK and Corradini representations, and the one point where the MCS data and the KK representation of the VDMC data disagree by a factor of two in the removed fraction, at $2.01k_F$ and $r_s = 2$, are collected in Appendix B. The statistical errors on the removed fractions are about two percentage points. The spread between data sets and between representations is the systematic uncertainty and is larger.

Figure 4(b) shows the spin factor $G_-$ at the same two densities. Its error bars are smaller than those of $G_+$, the KK curves pass through every point, and $G_-$ is smooth through $2k_F$. Both measured factors are therefore accurate, mutually consistent, and free of significant $2k_F$ structure, which is the accuracy this section needs.

Figure 4(c) is the spin-resolved subtraction on the data points themselves. The antiparallel factor rises smoothly and shows nothing at $2k_F$. The parallel-spin correlation is zero to about $0.7k_F$, then digs a deep minimum that reaches about $-1$ at $1.953k_F$ and closes to about $-0.2$ one grid step later. That is the mirror of the exchange peak, seen directly in the data.

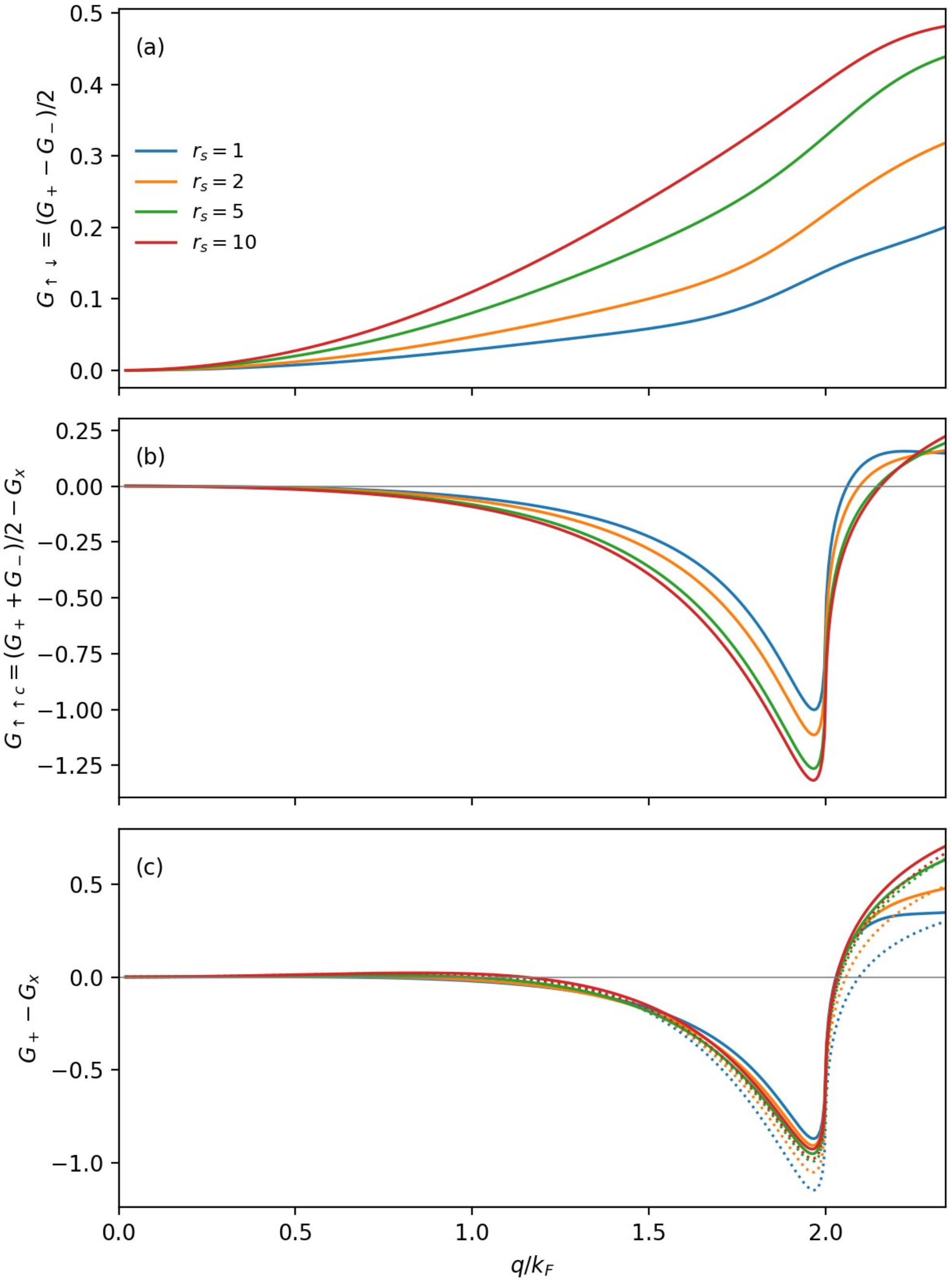

Fig. 5. Spin-resolved decomposition of the Kaplan-Kukkonen local field factors [6] about the exact exchange curve, Eq. (6), at $r_s$ = 1, 2, 5, and 10. (a) The antiparallel factor $G_{\uparrow\downarrow} = (G_+ - G_-)/2$. (b) The parallel-spin correlation $G_{\uparrow\uparrow c} = (G_+ + G_-)/2 - G_x$. (c) Their sum $G_+ - G_x$. Dotted curves in (c) are the same quantity from the Corradini representation [7]. Here $G_{\uparrow\uparrow}$ and $G_{\uparrow\downarrow}$ are half-normalized so that $G_\pm = G_{\uparrow\uparrow} \pm G_{\uparrow\downarrow}$. The rMCP07 paper of Ref. [45] defines $G_\pm$ with an additional factor 1/2. The KK forms are fitted for $q \leq 2.34k_F$.

Figure 5 shows the same decomposition through the KK forms at the four densities where $G_+$ is anchored to data. The contrast between the two channels is the central result of this paper. The argument has seven steps.

First, exchange is exact, universal, and peaked at $2k_F$. It enters $G_+$ and $G_-$ identically because the Pauli principle acts only between like spins.

Second, $G_+ - G_- = 2G_{\uparrow\downarrow}$ therefore contains no exchange. The antiparallel channel cannot cancel exchange structure. Whatever removes the peak from $G_+$ is parallel-spin, by identity.

Third, the Monte Carlo points show the peak removed by roughly half, and $G_+$ smooth through $2k_F$ within resolution. That is the "true" density local field factor.

Fourth, the measured $G_-$ is also smooth through $2k_F$, with smaller error bars still. Exchange enters $G_-$ with the same weight as $G_+$, so the same parallel-spin removal must operate in the spin channel, and it does.

Fifth, the parallel-spin correlation in Fig. 5(b) is accordingly large, negative, and sharply structured. Its minimum just below $2k_F$ mirrors the exchange maximum and removes nearly half of it. Within a smooth representation the mirror in shape is an identity. Its depth is what the data fix.

Sixth, the antiparallel factor in Fig. 5(a) rises smoothly and shows no feature at $2k_F$. This is what a Coulomb-driven effect looks like. Opposite-spin electrons are uncorrelated in Hartree-Fock, and switching on the interaction digs a repulsion hole where none existed. That process contains no exchange line and cannot supply a cubed logarithm. It reaches the Fermi surface only through the particle-hole bubbles it is built from, which carry at most the ordinary Lindhard feature at $2k_F$. It is observed to be smooth, and that is what one would expect.

Seventh, neither channel has been calculated from first principles near $2k_F$. The parallel correlation must follow the geometry of the Fermi surface to do what it does. A candidate mechanism is named in Section VI.

Panel (c) is the sum, and it is what a calculation of $G_+$ actually sees. The two channels partly cancel below $2k_F$, leaving a remainder that is small and featureless out to about $k_F$ and then falls steeply to about $-0.9$ near the exchange peak at every density shown. The near-$q^2$ growth of $G_+$ below $2k_F$ is that partial cancellation, and the smoothness through $2k_F$ is the deeper one.

The residual near the maximum is nearly the same at every density shown, about $-0.9$ from $r_s = 1$ to $10$, while $G_x$ is fixed at 1.99. Yet the residual must vanish as $r_s \to 0$, because correlation is of higher order in $e^2$ than exchange. So the removed fraction has no small-$r_s$ expansion. There is a crossover density, defined as the $r_s$ below which correlation is too weak to remove the exchange maximum, so that exchange-driven structure would survive in $G_+$. The maximum is already destroyed at some $r_s$ well below one, and every metallic density lies on the far side of the crossover. A rough scale comes from the width of the maximum. The Thomas-Fermi wavevector $k_{TF}/k_F = 0.81r_s^{1/2}$ equals a tenth of $k_F$ at $r_s \approx 0.015$ and equals the $0.46k_F$ width of the window in which $G_x > 1$ at $r_s \approx 0.3$. The crossover lies in that range. The fraction removed thus measures how far the metallic gas sits beyond the crossover rather than the strength of correlation. It points to screening of the exchange line as the mechanism, which Section VI takes up. The KK forms, fitted for $r_s \geq 1$, do not encode this limit. The statement rests on the exact limit and not on the fits.

The parallel channel also carries the known large-q behavior, the $C(r_s)(q/k_F)^2$ growth of Section II, which comes from the modification of the occupation numbers away from the noninteracting step when exchange and correlation are switched on. It is a single-particle effect that does not distinguish relative spins, so it enters $G_+$ and $G_-$ with the same coefficient and cancels from their difference. In Fig. 5 the antiparallel curves flatten while the parallel curves climb. The plateau values are fixed by the Zhu-Overhauser relation in terms of the pair function at zero separation [24], and the conversion between reference-function conventions is given by Takada [46]. The antiparallel factor is thus free of exchange, of the occupancy correction, and of the convention ambiguity. It is pure two-body correlation between electrons of opposite spin.

Higher-order exchange is not separately known. It does not change the contact value or stabilize the uniform state, and under the present definition it is counted with correlation [10,40].

Three caveats apply. The KK representations are smooth fits, so they cannot resolve whether the true $G_+$ retains a weak non-analyticity at $2k_F$. MCS, finding no evidence of $2k_F$ structure, caution that they cannot conclusively rule it out [3]. The obstacle is resolution rather than accuracy. The VDMC grid spacing near $2k_F$ is $0.078k_F$, and across the single interval from 1.953 to $2.031k_F$ the exact $G_x$ falls from 1.977 to 1.120. The whole cusp sits between two adjacent data points. The Corradini representation, fitted to MCS alone, is the natural sensitivity check and is shown in Fig. 5. Above $r_s = 5$ the $G_-$ entering the decomposition is the KK extrapolation rather than a fit to data. Finally, the decomposition of the interacting $G_{\uparrow\uparrow}$ into exchange and correlation is a definition, made unique here by taking the exchange part to be the exact first-order kernel.

## V. Benchmarks: the exact curve against the two published exchange-only kernels

This section uses Eq. (6) as a benchmark. Two published constructions cover the full dynamic exchange-only surface and reduce at $\omega = 0$ to a static kernel that can be compared with the exact curve point by point. The comparison checks Eq. (6) against independent numerical work and shows where each construction can be trusted.

There is no single object called the exact exchange kernel. There are three constructions but only two distinct objects, and the third is not obliged to agree with the first two. The first is the consistent first-order kernel, the $O(e^2)$ exchange contribution to the proper polarizability. It is the complete set of first-order exchange diagrams, two self-energy insertions and one vertex correction [17], it is unique, and it is Eq. (6). The second is the exact-exchange kernel of time-dependent density functional theory. For the uniform gas the two are the same object by construction, since the Kohn-Sham system is the free electron gas and the exact-exchange kernel is the $O(e^2)$ term of $f_{xc}$ in Görling-Levy perturbation theory [16]. Hellgren and von Barth showed the equivalence with time-dependent Hartree-Fock at first order [14], and Colonna, Hellgren, and de Gironcoli recovered the Engel-Vosko result in the static limit [15]. Their calculation is called RPAx, the random phase approximation with the exact-exchange kernel retained in the Dyson equation. The third is the McLachlan-variational kernel of Nazarov and Silkin (NS) [47,37], obtained by minimizing the mean-square residual of the time-dependent Schrödinger equation over Slater determinants. It is linear in the Coulomb interaction, so it is exactly first order in $e^2$ and shares the universality, but it is not built to reproduce the first-order diagram. The comparison measures how far one construction lies from the other. The constraints tested below follow from the Pauli principle and from exact sum rules and bind every construction.

The first construction is BDL's. They derived a dynamical-exchange-decoupling kernel from the equation of motion of the Wigner distribution function, with analytic reduction in 1976 and 1977 [28,13] and numerical tables in 1980 [30,20]. Holas, Aravind, and Singwi evaluated the first-order diagrams at arbitrary wavevector and frequency in the same period [11]. Hellgren and von Barth cite the BDL papers as prior derivations of the TDEXX response and state that "to first order in v, TDHF and TDEXX are the same" [14]. Table III gives the comparison, together with the 1977 tabulation of Tripathy and Mandal.

Table III. The static exchange-only local field factor. The exact expression, Eq. (6), against the Tripathy-Mandal 1977 tabulation [19], the BDL 1980 tabulation, $\nu = 0$ row of their Table I [30], and the Nazarov-Silkin McLVP kernel interpolated on the supplied $r_s = 2$ grid [37]. The large-q entries for TM and BDL are their analytic limits, not part of the tabulations.

| **$q/k_F$** | **Exact, Eq. (6)** | **Tripathy-Mandal 1977 [19]** | **BDL 1980 [30]** | **McLVP [37]** |
|---|---|---|---|---|
| 0.4 | 0.0409 | 0.0405 | 0.0405 | 0.0404 |
| 0.8 | 0.1757 | 0.1737 | 0.1761 | 0.1665 |
| 1.0 | 0.2907 | 0.2873 | 0.2898 | 0.2658 |
| 1.2 | 0.4513 | 0.4458 | 0.4492 | 0.3927 |
| 1.4 | 0.6773 | 0.6686 | 0.6733 | 0.5492 |
| 1.6 | 1.0047 | 0.9910 | 0.9924 | 0.7342 |
| 1.8 | 1.4988 | 1.4769 | 1.4620 | 0.9294 |
| 2.0 | $\pi^2/6 = 1.6449$ | 1.7511 | 1.5310 | 0.9459 |
| peak | 1.9924 at 1.9685 | 1.964 at 1.97 | not resolved, $0.2k_F$ grid | 1.013 at 1.94 |
| $q \to \infty$ | 1/3 | 1/3 (their Eq. 4.15) | 1/3 (Ref. [13]) | 1/2 |

The BDL $\nu = 0$ row lies on the exact curve to better than one percent for $q \leq 1.4k_F$, within 2.5 percent at $1.8k_F$, and 6.9 percent low at exactly $2k_F$, where their quadrature was not adapted to the singularity. Tripathy and Mandal, three years earlier, lie 0.9 to 1.5 percent below the exact curve at the tabulated wavevectors below $2k_F$ and 6.5 percent above it at $2k_F$ itself, for the same reason. BDL also verified that their kernel satisfies both the Niklasson and Kimball contact relations, obtaining $g(0) = 1/2$ from each [13]. More than forty years ago two groups had already tabulated what is here called the exact benchmark to about one percent, and both failed only at the one point where the analytic value is needed.

The NS McLVP kernel agrees with the exact curve at small q, where both are pinned by the compressibility sum rule. It falls progressively below it, by 8.5 percent at $k_F$ and by a factor of 1.74 at $2k_F$. Its smooth peak of 1.013 at $1.94k_F$ is half the exact peak, and it carries no counterpart of the cubed-logarithm cusp. The mean-square criterion is insensitive to a measure-zero non-analyticity, so it is not built to. Two framework-independent tests sort this. At large q the kernel tends to 1/2, the McLVP asymptote given by Nazarov [47, Eq. (21)]. Read through the Niklasson relation that would correspond to $g(0) = 1/4$, where the exchange-only value is 1/2 and the limit 1/3. The value 1/2 is also the large-q limit of the Hubbard factor [21], whose known defect is the implied contact probability. And at high frequency and long wavelength, writing $f_x(0,\omega\to\infty) = -c\pi/k_F^2$, McLVP gives $c = 3/4$ [47, Eq. (19)], while the third-frequency-moment sum rule with the exchange-only energy $\varepsilon_x = -3k_F/4\pi$ fixes $c = 3/5$ [1,48], as Nazarov himself notes. The two differ by a factor 5/4.

Two of the NS conclusions bear on the present result. From their own static comparison NS concluded that "the dip in question is due to exchange, not to correlations" [37], which partly anticipates the present decomposition. The exact benchmark adds the quantitative correction, since the exact exchange value near $2k_F$ is roughly twice the Monte Carlo value. And NS motivate McLVP by the dynamic pathologies of first-order theory, which are real [1] but do not bear on the static limit, where the first-order kernel is not an approximation but the unique exact $O(e^2)$ result. The practical conclusion is a domain of validity. The NS McLVP kernel covers the full dynamic surface on dense grids, which a time-dependent density functional calculation can use, and no published first-order tabulation offers that. Measured against the exact static result it is accurate to two percent below about $0.5k_F$ and to 8.5 percent at $k_F$. Quantities sensitive to the region $1.5k_F$ to $2k_F$, to the $2k_F$ non-analyticity, or to contact structure at large q should be referred to Eq. (6).

## VI. Conclusions

The exact static exchange-only local field factor of the uniform electron gas is written here as one expression, Eq. (6), assembled from known results. The function has been tabulated since 1977. What is new is the expression, its exact values, the Python module, and the use made of it here. It is one universal curve. It takes the value $\pi^2/6$ at exactly $2k_F$, where its slope diverges as a cubed logarithm, and it satisfies the compressibility sum rule and the exchange-only contact relation $g(0) = 1/2$. With it, exchange no longer needs to be computed, minimized, or fitted, and subtracting it from the Monte Carlo local field factors leaves correlation alone.

Three consequences follow. First, exchange and parallel-spin correlation near $2k_F$ are a nearly cancelling pair. The parallel correlation is large, negative, and peaked where exchange is peaked, removing roughly half of the exact exchange value at $r_s = 1$ to 10, while the antiparallel channel is small and featureless within the resolution of the data. Only the sum is smooth. The nearly quadratic $G_+$ that underlies the flat ALDA kernel $f_{xc} = -4\pi e^2 G_+/q^2$ is a consequence of that cancellation, not of any smoothness in exchange itself. The exchange kernel alone rises 65 percent from $q = 0$ to $2k_F$. Second, the stability of the metallic electron gas against the exchange-driven spin density wave at $r_s = 4.96$ and charge density wave at $r_s = 10.62$ is a parallel-spin effect of screening and correlation, everything beyond first-order exchange. Third, any kernel built on exact exchange must supply that correlation to be correct near $2k_F$. Correlation there is comparable in magnitude to the full local field factor and is not smooth, so it cannot be treated as a small correction to an exchange skeleton, and a correlation model lacking the $2k_F$ structure will reproduce the exchange-only charge density wave near $r_s \approx 10.6$.

Fermi-liquid theory expects the exact static response to carry only the ordinary Kohn anomaly at $2k_F$ [49,1], so the higher-order terms must collectively soften the cubed-logarithm slope of the first-order kernel. A parallel-spin mirror of the exchange structure is expected in form. What no argument of that kind predicts is its magnitude, and that is what the Monte Carlo data measure. Whether the true $G_+$ is smooth through $2k_F$ or retains a weak remnant is beyond the resolution of the present Monte Carlo data, whose grid spacing near $2k_F$ is $0.078k_F$, Section IV. Either way the parallel correlation must follow the geometry of the Fermi surface to do this, and I am not aware of a first-principles calculation of it near $2k_F$.

As a benchmark, the exact result sorts the existing exchange-only constructions. The 1977 Tripathy-Mandal and 1980 BDL tabulations sit within one to 2.5 percent of the exact curve below $2k_F$ and fail only at $2k_F$ itself [19,30]. The NS McLachlan-variational kernel [47,37] departs from it near $2k_F$ and does not meet the exchange-only contact and third-moment constraints.

One caution belongs here. Everything above concerns the uniform electron gas. Charge density waves in real materials arise from Fermi-surface nesting, from coupling to phonons, or from reduced dimensionality. The uniform gas has none of these. The statement here is narrower and, within its scope, sharp. In the uniform electron gas neither the spin density wave nor the charge density wave that exchange alone predicts occurs at all, because correlation removes about half of the exchange peak that would drive them.

Opportunities for further research follow directly from the decomposition. Three are listed here. The first is ours, the second is for the Monte Carlo community, and the third is the companion paper.

The parallel-spin correlation has a natural candidate mechanism, and it is the subject of our next paper. Engel and Vosko showed that the cubed logarithm arises in the vertex graph from the unscreened Coulomb line. Screening that line is parallel-spin correlation under the present definition. It cuts off precisely the small-momentum region that produces the singularity and leaves the large-q limit untouched. The crossover density defined in Section IV sits where the screening length first reaches the width of the exchange maximum, at $r_s$ well below one. That is the scale a screened exchange line would set. The leading antiparallel term is second order in the interaction and cannot be defined without screening, and it will be examined in the same place.

The decomposition also suggests specific Monte Carlo calculations for others. A finer grid near $2k_F$, of order $0.01k_F$ against the present $0.078k_F$, would establish whether the cancellation is exact or leaves a remnant. What is needed is resolution rather than smaller error bars. Computing $G_+$ by VDMC at $r_s$ = 3, 4, and 5, where $G_-$ already exists on that grid, would give both spin channels directly, and $G_-$ at $r_s$ = 10 would anchor the antiparallel channel where $G_+$ already is. A first-principles calculation of the antiparallel correlation alone would test the claim that it is the simpler object.

The dynamic case is exactly defined but has no explicit expression, and it will be treated in a companion paper. The analytic dynamic kernels in current use, CP07 and the MCP07 family [45], and the kernel of Panholzer, Gatti, and Reining [50], interpolate exchange and correlation together. The exact exchange surface separates the two. The dynamic kernel is also now being computed directly by diagrammatic Monte Carlo [51].

## Appendix A. The Engel-Vosko function I(Q)

With $l(x) = \ln|(1 + x)/(1 - x)|$, Eq. (29) of Ref. [17] reads

$$I(Q) = -[(1 - Q^4)/48Q^3]\, l(Q)^3 + [(1 - Q^2)/24Q^2]\, J_3(Q) - (1/8)[1/Q + ((1 - Q^2)/2Q^2)\, l(Q)]\, J_2(Q), \qquad \text{(A1)}$$

$$J_n(Q) = \int_0^Q dx\, [(1 - x^2)/x^2]\, l(x)^n, \quad n = 2, 3. \qquad \text{(A2)}$$

Here $Q = q/2k_F$ as in the text and x is the integration variable. The form is explicit in elementary functions plus the two one-dimensional integrals (A2). Fully integral-free forms exist. Glasser [18] gives I(Q) for $q \le 2k_F$ in logarithms, dilogarithms, and trilogarithms, his Eq. (19), together with the closed form of the n = 2 integral, his Eq. (22). Qian [12] gives a one-quadrature form, his Eq. (88), a series convergent at all q, his Eq. (93), and the expansion at $2k_F$, his Eq. (143). Special values are I(0) = −1, $I(1) = -\pi^2/24$ using $\int_0^1 dx\,(1 - x^2)x^{-2}\, l(x)^2 = \pi^2/3$ [17], the small-Q series $I(Q) = -1 + Q^2/9 + (46/675)Q^4 + \ldots$ whose leading coefficients were first obtained numerically by Kleinman and Lee [33] and by Chevary and Vosko [33], and the large-Q term $I(Q) \to -(1/27)Q^{-6}$, first obtained by Kleinman [35]. The derivative of I(Q) diverges as $(\ln|Q - 1|)^3$ at Q = 1 [17], and Qian's expansion $I(q \to 2k_F) = -\pi^2/24 + (1/12)(1 - q/2k_F) \ln^3|(1 - q/2k_F)/2|$ makes the cubed logarithm explicit [12, Eq. (143)]. Combined with Eq. (5) these give the results quoted in the text. Every number in this paper was evaluated from Eq. (A1) alone, with $J_2$ and $J_3$ by adaptive quadrature and the point x = 1 as an explicit breakpoint for Q > 1. That implementation, ev_gx.py, deposited in Ref. [38], is the sole numerical authority here. Qian proved Eq. (A1) rigorously and records two discrepancies with auxiliary expressions of Ref. [18], including Glasser's expansion at $2k_F$, while Glasser had earlier proved Eq. (A1) for $q \le 2k_F$. The correctness of Eq. (A1) rests on two independent proofs and on the numerical agreement. Qian's equation numbers follow paper II of Ref. [12], arXiv:1609.07741. The erratum to Ref. [17] concerns only the gradient-expansion coefficients of their Appendix E and touches nothing used here.

## Appendix B. Numerical values used in the text, with sources

Every number in the body is either exact, from ev_gx.py, the Python module of Section II [38], or in Table IV below with its source. Percentages are $(G_x - G_+)/G_x$ at the stated wavevector. Errors on Monte Carlo quantities are the published statistical errors, propagated. KK values use the AKCK_LFF package [6] with q converted from $k_F$ units to $\text{bohr}^{-1}$.

Table IV. Numerical values used in the text, with sources.

| **Quantity** | **Value** | **Where** | **Source** |
|---|---|---|---|
| $G_x$ at the VDMC point nearest the maximum | 1.977 | $q = 1.953k_F$ | ev_gx.py |
| VDMC $G_+$ there | 1.138 ± 0.036, 1.061 ± 0.037 | $r_s$ = 1, 2 | Ref. [4,5] data, K. Chen |
| Fraction of exchange removed there | 42.4 ± 1.8, 46.3 ± 1.9 percent | $r_s$ = 1, 2 | $(G_x - G_+)/G_x$ |
| $G_x$ at the MCS point nearest the maximum | 1.529 | $q = 1.81k_F$ | ev_gx.py |
| MCS $G_+$ there | 0.84(1), 0.88(2), 0.92(1) | $r_s$ = 2, 5, 10 | Ref. [3], Table I |
| Fraction removed there | 45.1 ± 0.7, 42.5 ± 1.3, 39.8 ± 0.7 percent | $r_s$ = 2, 5, 10 | |
| $G_x$ one VDMC grid step above $2k_F$ | 1.120 | $q = 2.031k_F$ | ev_gx.py |
| VDMC $G_+$ there | 1.097 ± 0.044, 1.112 ± 0.108 | $r_s$ = 1, 2 | equal to $G_x$ within errors |
| Parallel-spin | −0.97, −1.10 at | $r_s$ = 1, 2 | Fig. 4(c) |
| correlation from VDMC points, $(G_+ + G_-)/2 - G_x$ | $1.953k_F$. −0.17, −0.24 at $2.031k_F$ | | |
| The one MCS point above $2k_F$ | $G_+ = 0.90(7)$ at $2.01k_F$, $G_x = 1.328$, 32 ± 5 percent removed | $r_s = 2$ | Ref. [3] |
| KK at the same point | $G_+ = 1.106$, 17 percent removed | $r_s = 2$ | AKCK_LFF, the disagreement discussed in the text |
| Fraction removed at the exact maximum through KK | 43.7, 45.5, 47.7, 46.4 percent | $r_s$ = 1, 2, 5, 10 at $1.9685k_F$ | AKCK_LFF |
| Same through Corradini | 57.6, 52.8, 49.8, 49.0 percent | $r_s$ = 1, 2, 5, 10 | AKCK_LFF g_corradini; $r_s = 1$ lies outside the MCS anchor densities |
| Fraction removed at exactly $2k_F$ through KK | 31, 33, 35, 33 percent | $r_s$ = 1, 2, 5, 10 | AKCK_LFF |
| Exact maximum | 1.9924 at $1.9685k_F$ | | ev_gx.py, bounded maximization plus scan |
| Window in which $G_x > 1$ | $1.598k_F < q < 2.054k_F$ | | ev_gx.py |
| SDW threshold minimum | $r_s = 4.965$ at $1.832k_F$ | | Eq. (11), ev_gx.py |
| SDW threshold at $q \to 0$ and at $2k_F$ | $\pi\alpha$ = 6.029, 7.33 | | Eq. (11) |
| CDW threshold minimum | $r_s = 10.620$ at $1.944k_F$, where $G_x = 1.957$ | | Eq. (11) |
| CDW threshold elsewhere | 18.7 at $2k_F$, 52 at $1.65k_F$, 1100 at $1.60k_F$ | | Eq. (11) |
| Bloch ferromagnetic transition, for comparison | $r_s = 5.45$ | $q = 0$ | Ref. [39] |
| BDL 1980 against exact | 0.2 to 1.0 percent below $1.4k_F$, 1.2 at 1.6, 2.5 at 1.8, 6.9 at $2.0k_F$ | | Table III |
| Tripathy-Mandal 1977 against exact | 0.9 to 1.5 percent low below $2k_F$, 0.05 percent above, 6.5 percent high at $2k_F$ | | Table III |
| McLVP against exact | 2 percent low below $0.5k_F$, 5 at $0.8k_F$, 8.5 at $k_F$, 13 at $1.2k_F$, factor 1.739 at $2k_F$ | $r_s = 2$ grid | Table III |
| Thomas-Fermi scale | $k_{TF}/k_F = 0.81\, r_s^{1/2}$, equals $0.1k_F$ at $r_s$ = 0.015 and $0.46k_F$ at $r_s$ = 0.3 | | $(k_{TF}/k_F)^2 = 4r_s/\pi\alpha$ |
| Rise of $G_x/(q/k_F)^2$ from 0 to $2k_F$ | 1/4 to $\pi^2/24$, a factor $\pi^2/6 = 1.645$ | | Eq. (6) |

## Data and code availability

The Python module ev_gx.py, which returns $G_x(q,0)$, I(Q), and L(Q) for any q and is verified against all analytic checks of Ref. [17], together with the scripts producing Figs. 1 to 5 and the threshold curves of Eq. (11), is openly available in a Zenodo repository, Ref. [38]. An earlier version of this work is available as a preprint [52]. The Nazarov-Silkin kernel data were provided by V. Nazarov [37]. The Tripathy-Mandal values in Table III and Fig. 2 are transcribed from Table I of Ref. [19]. The BDL values are the $\nu = 0$ row of Table I of Ref. [30]. The VDMC $G_+$ and $G_-$ points of Ref. [4,5] in Fig. 4 were provided by K. Chen. These third-party data are not included in the Zenodo deposit but are available from their original sources. The Kaplan-Kukkonen local field factors are available as the Python package AKCK_LFF, installable with pip install AKCK_LFF, described in Ref. [6]. Its wavevector argument is an absolute wavevector in atomic units, not $q/k_F$.

## Acknowledgments

I thank V. Nazarov for providing the numerical data of Ref. [37], his Fortran code, and helpful correspondence, and K. Chen for the VDMC data. AI tools (Anthropic Claude, Google Gemini, and OpenAI ChatGPT, 2026) assisted with reference cataloguing, Python code development, manuscript editing, figure preparation, and critical review. I produced the decomposition and drafted the paper before employing AI assistance. I have verified all scientific content, data, and code, and I am solely responsible for the paper.

## References

[1] G. Giuliani and G. Vignale, Quantum Theory of the Electron Liquid (Cambridge University Press, Cambridge, UK, 2005).

[2] C. A. Kukkonen and A. W. Overhauser, Electron-electron interaction in simple metals, Phys. Rev. B 20, 550 (1979).

[3] S. Moroni, D. M. Ceperley, and G. Senatore, Static response and local field factor of the electron gas, Phys. Rev. Lett. 75, 689 (1995).

[4] C. A. Kukkonen and K. Chen, Quantitative electron-electron interaction using local field factors from quantum Monte Carlo calculations, Phys. Rev. B 104, 195142 (2021).

[5] K. Chen and K. Haule, A combined variational and diagrammatic quantum Monte Carlo approach to the many-electron problem, Nat. Commun. 10, 3725 (2019).

[6] A. D. Kaplan and C. A. Kukkonen, QMC-consistent static spin and density local field factors for the uniform electron gas, Phys. Rev. B 107, L201120 (2023).

[7] M. Corradini, R. Del Sole, G. Onida, and M. Palummo, Analytical expressions for the local-field factor G(q) and the exchange-correlation kernel $K_{xc}(r)$ of the homogeneous electron gas, Phys. Rev. B 57, 14569 (1998).

[8] T. Dornheim, S. Groth, and M. Bonitz, The uniform electron gas at warm dense matter conditions, Phys. Rep. 744, 1 (2018); T. Dornheim, J. Vorberger, S. Groth, N. Hoffmann, Zh. A. Moldabekov, and M. Bonitz, The static local field correction of the warm dense electron gas: An ab initio path integral Monte Carlo study and machine learning representation, J. Chem. Phys. 151, 194104 (2019).

[9] N. D. Drummond, Z. Radnai, J. R. Trail, M. D. Towler, and R. J. Needs, Diffusion quantum Monte Carlo study of three-dimensional Wigner crystals, Phys. Rev. B 69, 085116 (2004).

[10] A. W. Overhauser, Spin density waves in an electron gas, Phys. Rev. 128, 1437 (1962); Exchange and correlation instabilities of simple metals, Phys. Rev. 167, 691 (1968).

[11] A. Holas, P. K. Aravind, and K. S. Singwi, Dynamic correlations in an electron gas. I. First-order perturbation theory, Phys. Rev. B 20, 4912 (1979).

[12] Z. Qian, Static dielectric function with exact exchange contribution in the electron liquid, J. Math. Phys. 56, 111901 (2015), arXiv:1506.08342; Dielectric function with exact exchange contribution in the electron liquid. II. Analytical expression, arXiv:1609.07741 (unpublished).

[13] F. Brosens, J. T. Devreese, and L. F. Lemmens, Frequency-dependent exchange correction to the dielectric function of the electron gas (II), Phys. Status Solidi B 80, 99 (1977); F. Brosens, L. F. Lemmens, and J. T. Devreese, The pair correlation function and the dielectric response in the homogeneous electron gas, Phys. Status Solidi B 82, 117 (1977).

[14] M. Hellgren and U. von Barth, Linear density response function within the time-dependent exact-exchange approximation, Phys. Rev. B 78, 115107 (2008).

[15] N. Colonna, M. Hellgren, and S. de Gironcoli, Correlation energy within exact-exchange adiabatic connection fluctuation-dissipation theory: Systematic development and simple approximations, Phys. Rev. B 90, 125150 (2014).

[16] A. Görling, Exact exchange kernel for time-dependent density-functional theory, Int. J. Quantum Chem. 69, 265 (1998); Exact exchange-correlation kernel for dynamic response properties and excitation energies in density-functional theory, Phys. Rev. A 57, 3433 (1998).

[17] E. Engel and S. H. Vosko, Wave-vector dependence of the exchange contribution to the electron-gas response functions: An analytic derivation, Phys. Rev. B 42, 4940 (1990); Erratum: Phys. Rev. B 44, 1446 (1991).

[18] M. L. Glasser, Exchange corrections to the static Lindhard screening function, Phys. Rev. B 51, 7283 (1995).

[19] D. N. Tripathy and S. S. Mandal, Calculation of the dielectric function for an electron liquid, Phys. Rev. B 16, 231 (1977).

[20] J. T. Devreese, F. Brosens, and L. F. Lemmens, Dielectric function of the electron gas with dynamical-exchange decoupling. I. Analytical treatment, Phys. Rev. B 21, 1349 (1980).

[21] J. Hubbard, The description of collective motions in terms of many-body perturbation theory. II. The correlation energy of a free-electron gas, Proc. R. Soc. London A 243, 336 (1957).

[22] D. J. W. Geldart and R. Taylor, Wave-number dependence of the static screening function of an interacting electron gas. I. Lowest-order Hartree-Fock corrections, Can. J. Phys. 48, 155 (1970); II. Higher-order exchange and correlation effects, 48, 167 (1970).

[23] D. J. W. Geldart, T. G. Richard, and M. Rasolt, Comments on the dielectric function of Toigo and Woodruff, Phys. Rev. B 5, 2740 (1972).

[24] X. Zhu and A. W. Overhauser, Exact exchange and correlation corrections for large wave vectors, Phys. Rev. B 30, 3158 (1984).

[25] F. Toigo and T. O. Woodruff, Calculation of the dielectric function for a degenerate electron gas with interactions. I. Static limit, Phys. Rev. B 2, 3958 (1970).

[26] F. Toigo and T. O. Woodruff, Calculation of the dielectric function for a degenerate electron gas with interactions. II. Frequency dependence, Phys. Rev. B 4, 4312 (1971).

[27] A. K. Rajagopal, Variational solution of vertex equation and dielectric function of an interacting-electron gas, Phys. Rev. A 6, 1239 (1972).

[28] F. Brosens, L. F. Lemmens, and J. T. Devreese, Frequency-dependent exchange correction to the dielectric function of the electron gas (I), Phys. Status Solidi B 74, 45 (1976).

[29] F. Brosens, L. F. Lemmens, and J. T. Devreese, Phys. Status Solidi B 81, 551 (1977), as cited in Ref. [30]; this paper was not available to the author.

[30] F. Brosens, J. T. Devreese, and L. F. Lemmens, Dielectric function of the electron gas with dynamical-exchange decoupling. II. Discussion and results, Phys. Rev. B 21, 1363 (1980).

[31] G. Niklasson, Dielectric function of the uniform electron gas for large frequencies or wave vectors, Phys. Rev. B 10, 3052 (1974).

[32] J. C. Kimball, Short-range correlations and electron-gas response functions, Phys. Rev. A 7, 1648 (1973).

[33] L. Kleinman and S. Lee, Gradient expansion of the exchange-energy density functional: Effect of taking limits in the wrong order, Phys. Rev. B 37, 4634 (1988); J. A. Chevary and S. H. Vosko, Importance of Pauli-principle restrictions for accurate numerical evaluation of the exchange-only screening function, Phys. Rev. B 42, 5320 (1990).

[34] P. S. Svendsen and U. von Barth, Gradient expansion of the exchange energy from second-order density response theory, Phys. Rev. B 54, 17402 (1996).

[35] L. Kleinman, Exchange density-functional gradient expansion, Phys. Rev. B 30, 2223 (1984).

[36] S. Ichimaru, Strongly coupled plasmas: high-density classical plasmas and degenerate electron liquids, Rev. Mod. Phys. 54, 1017 (1982).

[37] V. U. Nazarov and V. M. Silkin, Exchange kernel $f_x(q,\omega)$ of electron liquid from the variational principle of McLachlan, Phys. Rev. B 110, 104310 (2024).

[38] C. A. Kukkonen, Exact static exchange benchmark for the uniform electron gas, Zenodo (2026), doi:10.5281/zenodo.21497366.

[39] F. Bloch, Bemerkung zur Elektronentheorie des Ferromagnetismus und der elektrischen Leitfähigkeit, Z. Phys. 57, 545 (1929).

[40] L. Baguet, F. Delyon, B. Bernu, and M. Holzmann, Hartree-Fock ground state phase diagram of jellium, Phys. Rev. Lett. 111, 166402 (2013).

[41] N. Colonna, M. Hellgren, and S. de Gironcoli, Molecular bonding with the RPAx: From weak dispersion forces to strong correlation, Phys. Rev. B 93, 195108 (2016).

[42] M. Hellgren, N. Colonna, and S. de Gironcoli, Beyond the random phase approximation with a local exchange vertex, Phys. Rev. B 98, 045117 (2018).

[43] D. M. Ceperley and B. J. Alder, Ground state of the electron gas by a stochastic method, Phys. Rev. Lett. 45, 566 (1980).

[44] G. Vignale and K. S. Singwi, Effective two-body interaction in Coulomb Fermi liquids, Phys. Rev. B 32, 2156 (1985).

[45] L. A. Constantin and J. M. Pitarke, Simple dynamic exchange-correlation kernel of a uniform electron gas, Phys. Rev. B 75, 245127 (2007); A. Ruzsinszky, N. K. Nepal, J. M. Pitarke, and J. P. Perdew, Constraint-based wave vector and frequency dependent exchange-correlation kernel of the uniform electron gas, Phys. Rev. B 101, 245135 (2020);

A. D. Kaplan, N. K. Nepal, A. Ruzsinszky, P. Ballone, and J. P. Perdew, First-principles wave-vector- and frequency-dependent exchange-correlation kernel for jellium at all densities, Phys. Rev. B 105, 035123 (2022).

[46] Y. Takada, Emergence of an excitonic collective mode in the dilute electron gas, Phys. Rev. B 94, 245106 (2016).

[47] V. U. Nazarov, Time-dependent effective potential and exchange kernel of homogeneous electron gas, Phys. Rev. B 87, 165125 (2013).

[48] N. Iwamoto and E. K. U. Gross, Correlation effects on the third-frequency-moment sum rule of electron liquids, Phys. Rev. B 35, 3003 (1987).

[49] W. Kohn, Image of the Fermi surface in the vibration spectrum of a metal, Phys. Rev. Lett. 2, 393 (1959); J. M. Luttinger, Fermi surface and some simple equilibrium properties of a system of interacting fermions, Phys. Rev. 119, 1153 (1960).

[50] M. Panholzer, M. Gatti, and L. Reining, Nonlocal and nonadiabatic effects in the charge-density response of solids: A time-dependent density-functional approach, Phys. Rev. Lett. 120, 166402 (2018).

[51] J. P. F. LeBlanc, K. Chen, K. Haule, N. V. Prokof'ev, and I. S. Tupitsyn, Dynamic response of an electron gas: Towards the exact exchange-correlation kernel, Phys. Rev. Lett. 129, 246401 (2022).

[52] C. A. Kukkonen, Exact static exchange benchmark for the uniform electron gas: Dynamic-kernel tests and exchange-correlation cancellation near $2k_F$, arXiv:2607.20403 (2026).